\documentclass[epj]{svjour}
\usepackage{graphics}
\begin{document}
\title{Kinetics of water flow through polymer gel}
\author{Yasuo Y. Suzuki\inst{1,2}, Masayuki Tokita\inst{3} \and Sada-atsu Mukai\inst{3}
}                     
%
%
\institute{Institut de Physique Th\'{e}orique, CEA, IPhT, CNRS, URA 2306, F-91191 Gif-sur-Yvette, France, \and
Faculty of Engineering, Takushoku University, Hachioji, Tokyo 193-0985, Japan, \email{y-suzuki@la.takushoku-u.ac.jp}
\and Department of Physics, Faculty of Science, Kyushu University,  4-2-1 Ropponmatsu, Fukuoka 810-8560,  Japan}
\date{Received: date / Revised version: date}
%
\abstract{
The water flow through the poly(acrylamide)
gel under a constant water pressure is measured by newly designed apparatus. 
The time evolution of the water flow in the gel, 
is calculated based on the collective diffusion model of the polymer network 
coupled with the friction between the polymer network and the water. 
The friction coefficient are determined from the equilibrium velocity of water flow. 
The Young modulus and the Poisson's ratio of the rod shape gels are measured by the uni-axial elongation experiments, which determine
the longitudinal modulus independently from the water flow experiments.
With the values of the longitudinal modulus and of the friction determined by the experiments, 
the calculated results are compared with the time evolution of the flow experiments.  
We find that the time evolution of the water flow is well described by a single characteristic 
relaxation time predicted by the collective diffusion model coupled with the water friction.
} 
\maketitle
\section{Introduction}
\label{intro}

The gel is an important state of matter that is found in a wide
variety of biological, chemical, and food systems \cite{TanakaSci}.  The polymer
gel consists of a cross-linked polymer network and a large amount of
solvent (typically water).  Since the average mesh size (the distance between neighboring
crosslinks) of the polymer network is in general large compared to
small molecules, they can pass through the gel easily.  Two
different transport processes are important in the gel.

The first one is the diffusive flow of molecules \cite{Muhr}.
The transport of molecules by the diffusion in the polymer
network is modeled as the diffusion of the probe objects in
the fixed mesh of obstacles.  The ratio between the probe size and
the mesh size plays essential role to determine
the diffusion coefficient of the molecules in the gels.  
The diffusion of the probe molecules in the
gel has been measured by the pulsed field gradient nuclear magnetic resonance.  
The results indicate that the diffusion coefficient of the probe
molecules of various sizes in the gel is well described by a simple
scaling relationship \cite{TokitaPRE}.

The second process is the
convective flow of the water through the polymer network \cite{TokitaJCP95,TokitaSci,TokitaAPS}.  When the water flows in the
gel, it experiences the hydrodynamic friction from the polymer
network, at the same time, the polymer network of gel is deformed
from the initial configuration by the drag force of the
water.  The entire flowing process of the water is determined by the
balance between the viscoelastic response of the water flow and that of the polymer
network.  The collective diffusion of the polymer network coupled with
water friction,
therefore, plays essential roles in the convective flow
process in the gel.  

As well as the transport phenomena in the gel, the water flow 
is also important in the kinetics of the volume phase
transition of the gel including the pattern formations in shrinking gels\cite{TanakaPRL,TanakaJCP70,Matsuo,TokitaJPSJ,MaskawaJCP,TakigawaJCP111,TokitaJCP113,TakigawaJCP117,BoudaoudPRE,UrayamaJCP,NosakaPoly}.  
Kinetics of solvent flow in the gel is the main subject
of the application of the gel to control the solvent flow \cite{YoshikawaJJAP,SuzukiJCP}.  
Although the significance, the solvent flow process
in the polymer gel has yet to be studied in detail.

It is the purpose of this paper to describe the kinetics of the water
flow through the polymer gel, based on the collective diffusion model coupled with water friction.
The characteristic relaxation time of the water flow process in the gel is
expressed as $\tau = L^2/ \pi^2 D_c$ with the parameters: the typical size of the
gel, $L$, and the collective diffusion coefficient of the polymer
network, $D_c$.  The collective
diffusion coefficient is expressed by using the longitudinal modulus of the
gel, $\kappa$, and the friction coefficient between the
polymer network and the water, $f$, as \cite{TanakaJCP70,TanakaJCP59}
\begin{equation}
D_c = \frac{\kappa}{f}.
\label{dc}
\end{equation}
The water flow in
the polymer network is, thus, directly governed by the viscoelastic
property of the gel. 
The collective diffusion coefficient of polymer gels has been extensively 
measured by quasi-elastic light scattering \cite{Munch1,Munch2}.
The friction coefficient
between the polymer network of gel and the water has been measured
by flowing the water through the gel \cite{TokitaJCP95,TokitaSci,TokitaAPS}. 

In this paper, we present experimental results on the water
flow through the gel, and show that the time evolution of the water flow 
through the gel is described satisfactorily by the collective diffusion model of polymer network with water friction.
The information on the kinetics of water flow in the gel will help the
better use of them in the applications.

\section{Experiment}
\label{sec:2}
\subsection{Water flow measurements}
\label{subsec:2.1}

The experimental setting of the water flow measurements is illustrated schematically in Figure 1.
%
\begin{figure}
\resizebox{0.75\columnwidth}{!}{%
  \includegraphics{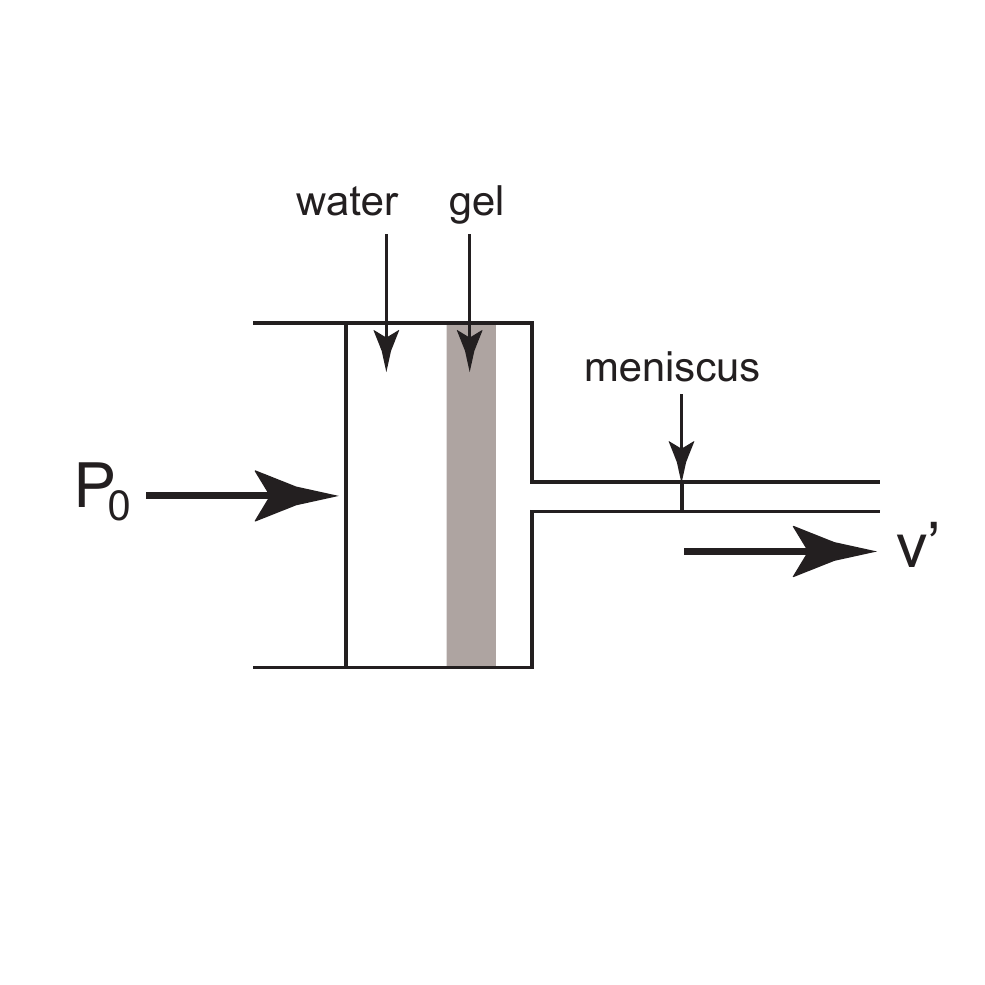}
}
\caption{Experimental setting of water flow measurements}
\label{fig:1}       
\end{figure}
A thin circular slab gel of thickness, $L$, is set in a tube and the rim of the gel is glued to the tube.  
The surface of the gel at the lower reaches side is
mechanically fixed to the end plate through an O-ring.  When the small pressure, $P_0$, is applied
to the water from the upper stream side of the gel, the water 
flows through the gel deforming the polymer network of gel by the
drag force.  The macroscopic value of the friction coefficient of the gel, $f$, is determined
from the velocity of the water flows out of the gel in the equilibrium
state, $v_{\infty}$, as
\begin{equation}
 f = \frac {P_0}{v_{\infty}L}.
 \label{friction}
\end{equation}

Even though the concept is simple, the mechanical measurement of the
friction coefficient between the polymer network of gel and the water is not easy
because the friction of the gel is huge. 
The difficulty of evaluating the friction coefficient
measured from the water flow through the gel is described in the reference \cite{TokitaJCP95}.  
In our experiments, the difficulty is overcome by using a well
calibrated homogeneous glass capillary to amplify the velocity of
water as shown in Figure 1.  The position of the meniscus of water in
the capillary is measured as a function of time after the pressure is
applied to the water.

The photographs of the apparatus used in this study are given in
Figure 2.  
%
\begin{figure}
\resizebox{0.75\columnwidth}{!}{%
  \includegraphics{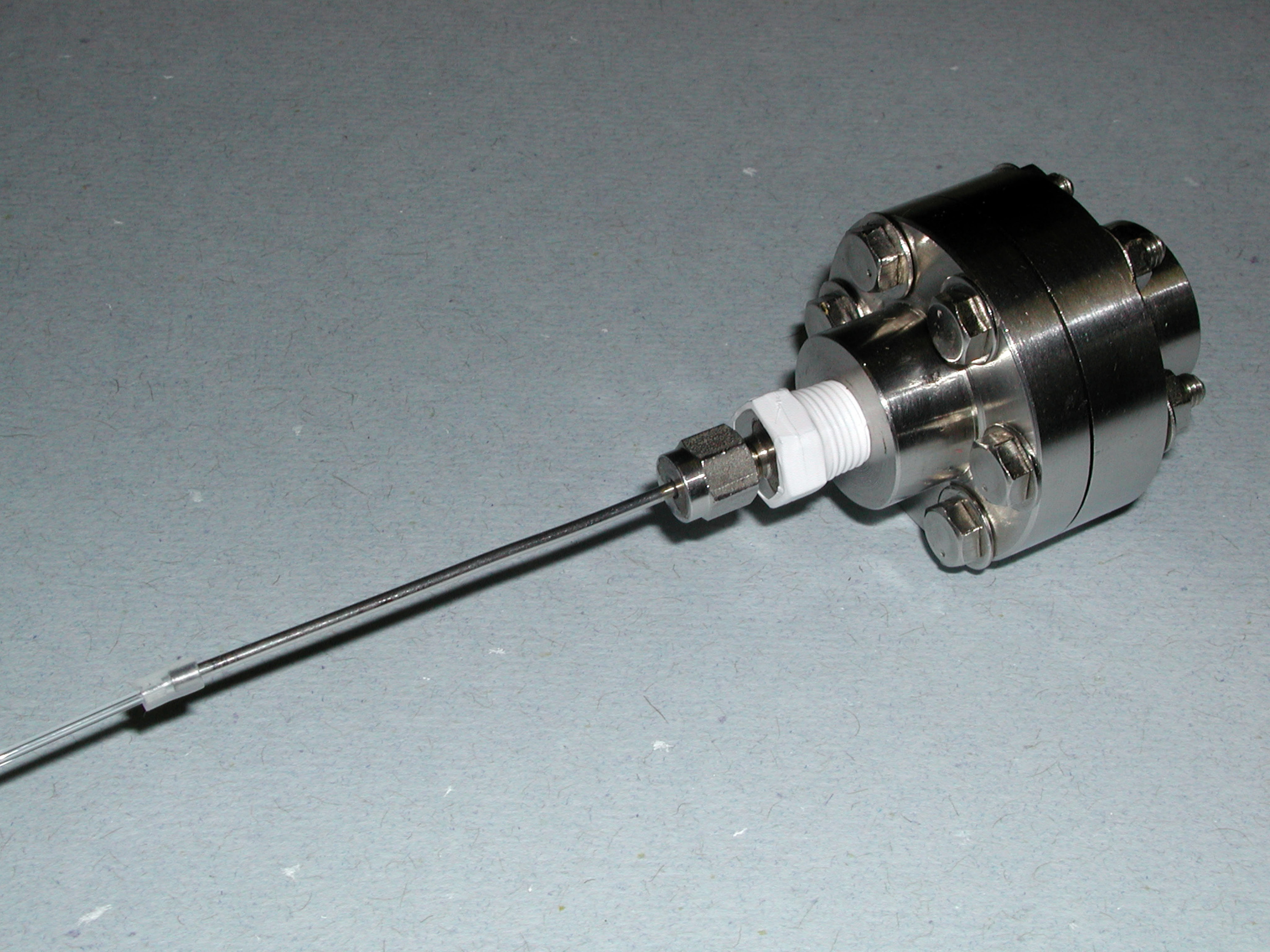}}
\resizebox {0.75\columnwidth}{!}{
  \includegraphics{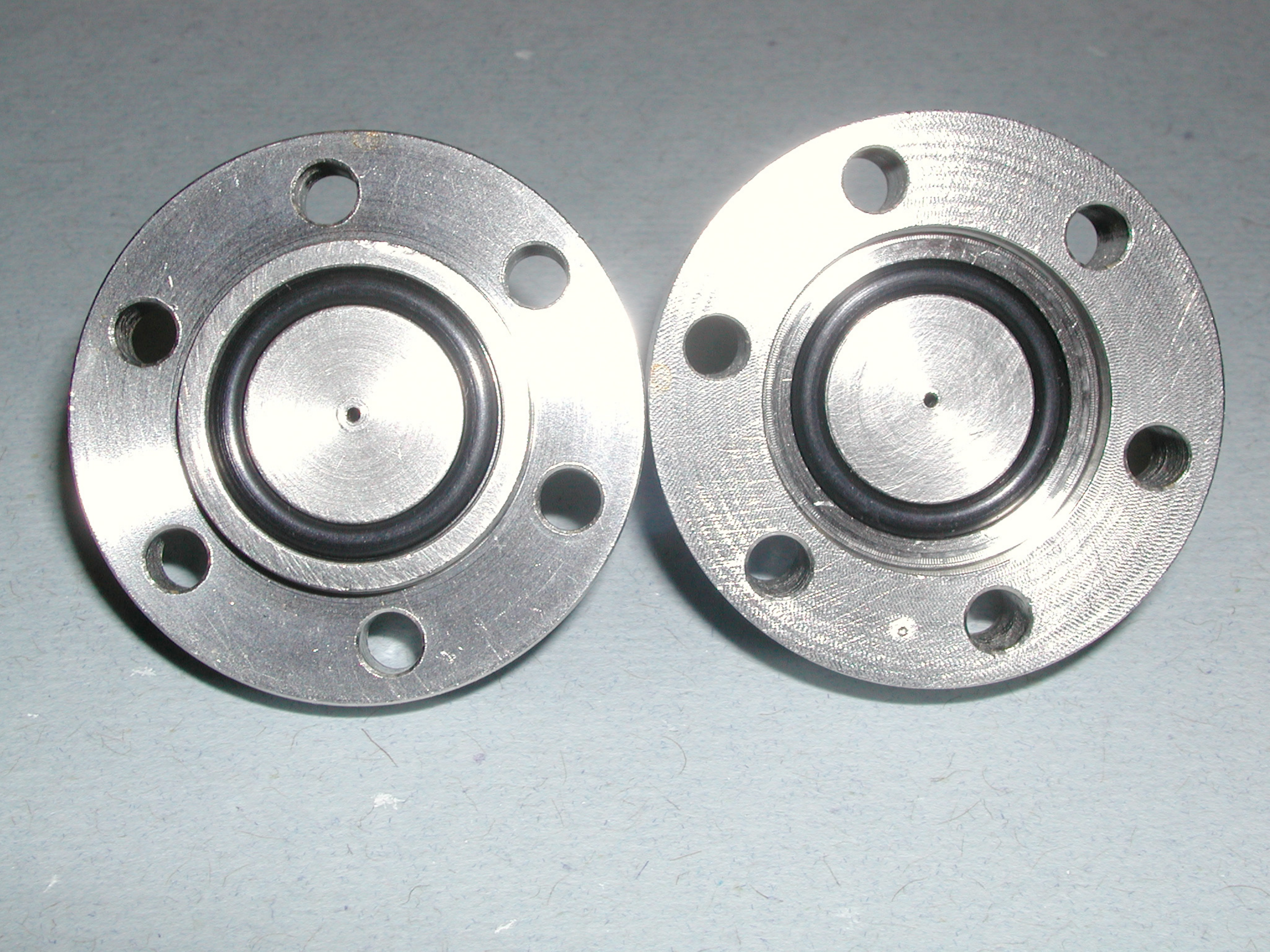}
}
\caption{Apparatus for water flow measurements: (a) the upper photograph shows the external appearance, (b) the lower photographs shows the inside of the end plates.}
\label{fig:2}       
\end{figure}
The essential structure of the apparatus is the same with
the previous one but some points are improved for the precise
measurements of the friction of the gel \cite{TokitaJCP95}.  The right hand side of
Figure 2 (a) is the upper stream side and is connected to a water
column to apply the hydrostatic pressure.  The glass capillary is set
at the top of stainless steel pipe which can be seen at the left hand
side of Figure 2 (a).  Leak of water around the gel is prevented
by setting O-rings inside the cell as shown in Figure 2 (b).  All
parts of the cell is made of stainless steel to avoid the mechanical
deformation due to the applied pressure.  The mechanical deformation
of the apparatus is considerably reduced compared to
the previous one.

\subsection{Mechanical response measurements}

The mechanical response of the polymer gel relates
the friction coefficient to the collective diffusion coefficient of the polymer network
as $D_{c}= \kappa /f$ where $\kappa= K+ \frac{4\mu}{3}$ \cite{TanakaJCP59}.
Here, $\kappa$ is the longitudinal modulus of the gel 
and $K$ and $\mu$ are the bulk modulus and the shear modulus
of the gel, respectively. 
We measure two elastic moduli
of the gel simultaneously, and the longitudinal modulus of the
gel is determined from them.

The rod shape gel is elongated in water at a strain of about 10 $\%$
and the stress is measured.  The size of the gel is observed by using a microscope
during the elasticity measurements to obtain the Young's modulus of
the gel, $E$, and the Poisson's ratio, $\sigma$.
The longitudinal modulus of the gel, $\kappa$, is determined
from the Young's modulus and the Poisson's ratio by $\kappa= K+ \frac{4\mu}{3}$, 
$K = E/3(1-2\sigma)$ and $\mu = E/2(1+\sigma)$.

In this short time experiments, 
the water solvent neither goes out nor comes into the gels.
Thus we determine the mechanical response of the gel independently
from the friction experiments by the water flow.  
We use the results to analyze the kinetics of water flow through the polymer
network.  

\subsection{Sample}

The samples used in this paper are poly(acrylamide) gel that is obtained by the
radical co-polymerization of the main-chain component, acrylamide, and
the cross-linker, N,N'-methylene-bis-acrylamide.  The concentration of
the gel used in this measurement is 1 M.  Ammoniumpersulfate and
N,N,N',N'-tetramethyl-ethylenediamine are used as the initiator and
the accelerator.  All chemicals used here are of electrophoresis grade,
purchased from BioRad and used without further purification.
The gel for the
friction measurements is prepared in the circular gel mold, where the gel-bond films
(FMC) with circular opening are glued on the both sides.   For the friction coefficient measurements,
the gel is prepared in the measurement cell and kept under water at a constant room temperature for overnight to reach
the equilibrium state.

For the elastic modulus measurements, the rod shaped gels are prepared in the capillary.  
The desired amounts of the
main-chain component, the cross-linker, and the accelerator are
dissolved into the distilled and de-ionised water, which is prepared
by a Mili-Q system.  The pre-gel solution is de-gassed for 20 min, and
then the desired amount of the initiator is added to the solution to
initiate the reaction.  The sample gels are taken out of the reaction
bath and then washed by distilled and de-ionized water extensively and
used in each measurement.

\section{Experimental results}
\label{sec:results}
\subsection{Friction coefficient of gel}
The experimental results of the measurement of water flow through
the gel are shown in Figure 3.  
\begin{figure}
\resizebox{0.75\columnwidth}{!}{%
  \includegraphics{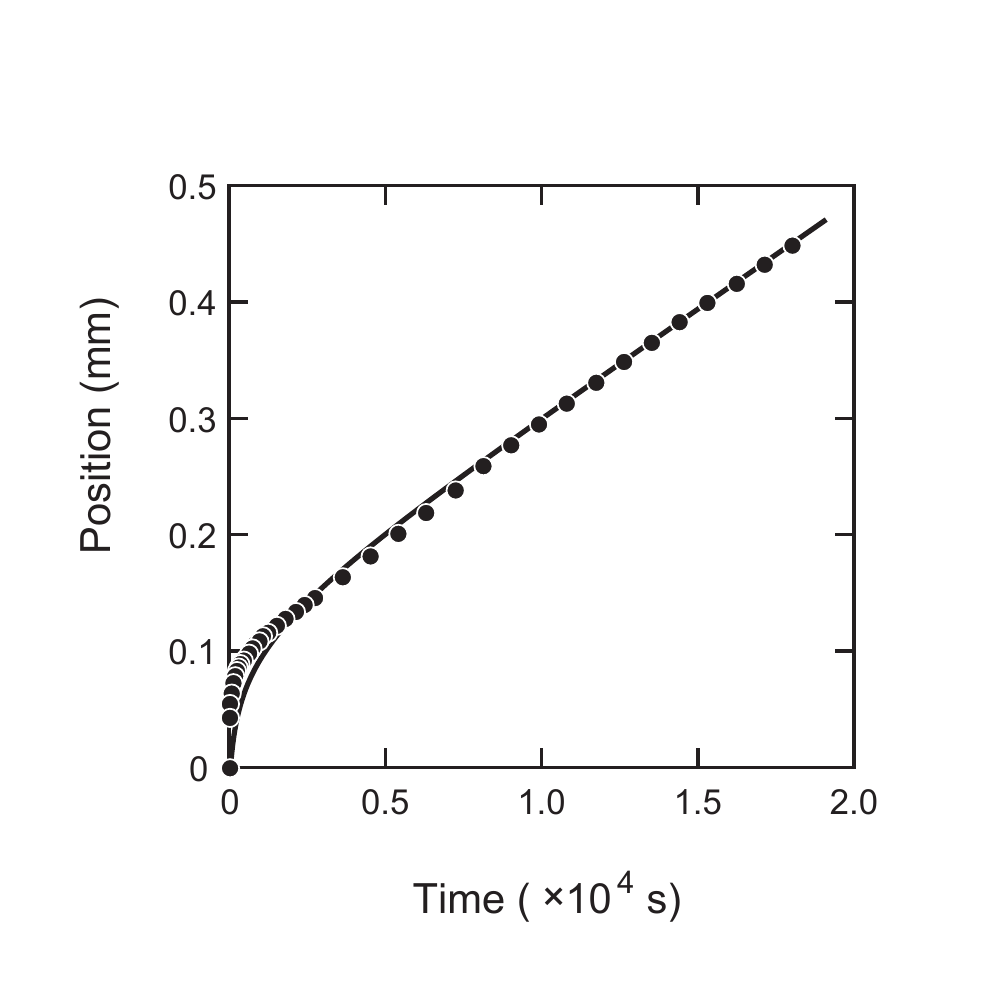}
}
\caption{Position of meniscus in capillary fit by equation (\ref{position}) with $a$ and $b$
as adjusted parameters.}
\label{fig:3}       
\end{figure}
The position of the meniscus in
the capillary of $10 \mu l$ glass micro-pipette is plotted as a
function of the time elapsed after the application of the pressure to
the water.  The slope of the flow curve, thus obtained, represents the
water velocity in the capillary, $v'$.  The
velocity of the flow is high at first.  Then it eventually decreases with
time and approaches to an equilibrium value.  The equilibrium state is
attained about 5 to 7 $\times$ 10$^3$ s after the pressure is applied
to the water.  The velocity of the meniscus in the equilibrium
state is determined by the least-squares analysis of the linear
portion of the flow curve in Figure 3, typically in the time region
more than 1 $\times$ 10$^{4}$ s, that yields to $v'_{\infty} = 1.9 \times
10^{-7}$ m/s.  The velocity of water flow in the gel, $v$, is then
calculated from the velocity of the meniscus, $v'$, with the ratio of the radius of the circular
opening of the gel mold, $R$, and that of the capillary, $r$, as $v =
v'(r/R)^2$.  The values of $R$ and $r$ are measured as 1.745 mm and
0.268 mm, respectively.  The equilibirium velocity of water flow in the gel is,
thus, calculated as $v_{\infty} = 4.4 \times 10^{-9}$ m/s.  The pressure
applied to the water is $P_0 = 2.9 \times 10^{3}$ N/m$^2$, which corresponds
to the height of water column of 30 cm, and the thickness of the gel
is $L = 1.0 \times 10^{-3}$ m.  From these values, the friction
coefficient of the gel is calculated by equation (\ref{friction}) that yields to $f
= 6.6 \times 10^{14}$ N s/m$^4$.  This agrees with a typical value of the
friction coefficient of the transparent poly(acrylamide) gel
evaluated by other experiments \cite{TokitaJCP95}.

\subsection{Elastic modulus of gel}
In Figure 4, we show the concentration dependence of the Young's
modulus and the Poisson's ratio of the gels used in our experiment.
\begin{figure}
\resizebox{0.75\columnwidth}{!}{%
  \includegraphics{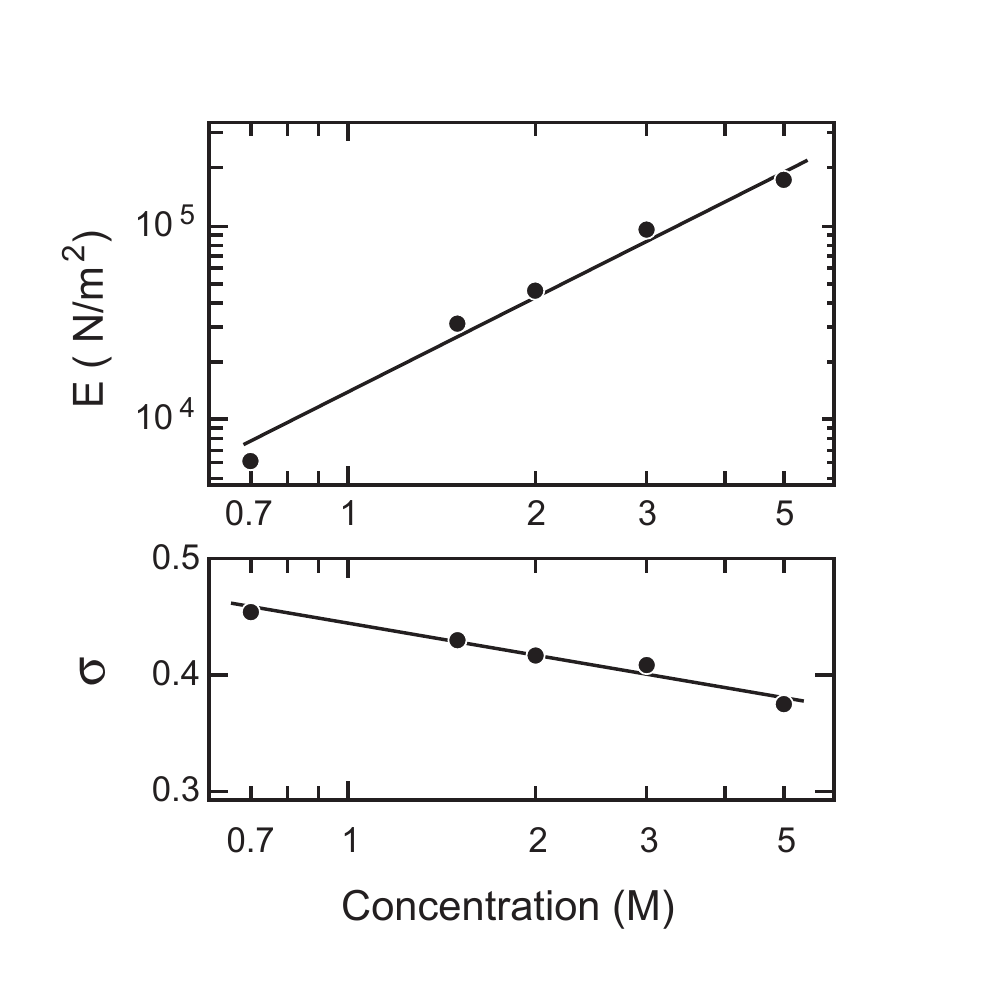}
}
\caption{Concentration dependence of elastic moduli: (a) the upper graph shows Young's modulus, $E$, (b) the lower graph shows Poisson's ratio, $\sigma$.}
\label{fig:4}       
\end{figure}
The least-squares analysis of the results yields that the
concentration dependence of the Young's modulus of the gel is well
described by a power law relationship with a scaling exponent of 1.7,
which is slightly smaller than the value expected from the scaling argument \cite{deGennes}.  
The Poisson's ratio is about 0.45 in
the dilute gels.  It decreases with the concentration of the gel and
reaches to about 0.35 at higher concentration region.  
The values of the Poisson's ratio of the dilute gels are close to those of the incompressible
materials.  On the other hand, 
the values of the Poisson's ratio of the dense gels are
rather close to those of the metals and the glasses.  
The Young's modulus and the Poisson's ratio of the gel
at a concentration of 1 M is observed as $E = 1.3 \times 10^{4}$
N/m$^{2}$ and $\sigma = 0.44$, respectively.  The longitudinal modulus
is then calculated as $\kappa = 4.1 \times 10^{4}$
N/m$^{2}$ for the gel at a concentration of 1 M.
Our results are consistent with previously published Poisson's ratio of 
poly(acrylamide) gels and a theory \cite{TakigawaPJ1,TakigawaPJ2}.

The diffusion coefficient of the polymer network is deduced from equation (\ref{dc}) as
$D_c =6.2 \times 10^{-11}$ m$^{2}$/s, which agrees with a typical value of the
diffusion coefficient of the transparent poly(acrylamide) gel evaluated by other
experiments \cite{TanakaJCP70,TanakaJCP59,Munch1,Munch2}.

\section {Kinetics of water flow through gel}

When the water flows in the gel (polymer network) by the application of the
mechanical pressure, the polymer network of gel deforms from the
initial position by the drag force of the water flow.  
We introduce a function, $u(x,t)$, that represents the
displacement of a point in the polymer network from its initial
position at $x$.  We assume $u$ is always small compared to the characteristic size of the gel, $L$.  
Under the present definition, $u(x, t) = 0$ at
$t=0$.   We also assume that the macroscopic velocity of the water in the polymer network, $v(t)$, and
microscopic friction coefficient, $f$,
are uniform in space. 
\begin{equation} 
 \frac{\partial v}{ \partial x }=0.
 \label{incomp}
\end{equation}
Then, we write down the equation of motion of the polymer network and that of the water in the
gel as
\begin {eqnarray}
\rho_g \ddot{u} &=& - f (\dot{u} - v) + \kappa \frac {\partial^2 u}{\partial x^2}, \label{mou}\\
\rho_w \dot{v} &=&  f(\dot{u} - v) - \frac {\partial P}{\partial x}, \label{mov}
\end {eqnarray}
where $\rho_g$ and $\rho_w$ are the density of the polymer network
and that of the water, respectively;   A dot over the variable denotes the time derivative;   $P(x,t)$ represents the pressure inside the region that  the gel occupied.

By neglecting the inertia terms in equations (\ref{mou}) and (\ref{mov}),
we obtain
\begin {equation}
 \kappa \frac{\partial^2u}{\partial x^2} - \frac {\partial P}{\partial x}=0,
\label{eqf}
\end {equation}
and
\begin{equation}
f(v-\dot{u})= - \frac {\partial P}{\partial x}.
\label{eqv}
\end{equation}
A set of equations (\ref{incomp}), (\ref{eqf}), (\ref{eqv}) corresponds to
one dimensional case under a limit of small polymer volume fraction $\phi \rightarrow 0$  of a general description for 3 dimensional deformation of gels\cite{Yamaue,Doi}.  

In our experimental setup of water flow measurement in Fig.\ref{fig:1}, 
we see $P(x=0,t)=P_0$ and $P(x=L,t)=0$.  
Taking a spacial average of equation (\ref{eqv}) over the space the gel occupied,
we obtain
\begin{equation}
 v-\frac{1}{L} \int_0^L \dot{u} dx = \frac{P_0}{Lf}.
 \label{P0}
\end{equation}
The time evolution of water flow in the gel $v(t)$ is determined locally from $u(x,t)$ by
\begin{equation}
v=\dot{u}-D_c\frac{\partial^2u}{\partial x^2}.
\end{equation}
Here, we use the collective diffusion coefficient of the gel, $D_c$, defined in equation (\ref{dc}). 
The displacement of polymer network $u(x,t)$ is, hence, determined by solving the equation:
\begin {equation}
 \dot{u}-\frac{1}{L}\int_0^L\dot{u}dx=D_c \frac{\partial^2u}{\partial x^2} + \frac{P_0}{Lf}.
\label{equ}
\end {equation}

In the equilibrium state, $\dot u(x, t) \rightarrow 0$ as $t \rightarrow \infty$.  
The restoring force due to the deformation of
the polymer network is in balance with constant gradient of the pressure.  
In the equilibrium state, therefore, equations (\ref{incomp}), (\ref{eqf}), (\ref{eqv}), (\ref{P0}) give
\begin {equation}
\frac {\partial^2 u(x, t= \infty )}{\partial x^2} = - \frac{P_0}{D_{c}Lf}.
\label{equilicond}
\end {equation}
The displacement, $u(x, t)$, satisfies the following boundary conditions:
\begin {eqnarray}
u(L,t) &=& 0,\\
\frac {\partial u(0, \infty)}{\partial x} &=& 0.
\end {eqnarray}
Integration of equation (\ref{equilicond}) with respect to $x$ with the boundary
conditons yields
\begin {equation}
u(x, \infty) =  \frac{P_0}{2D_cLf} (L^2 - x^2).
\label{equilibrium}
\end {equation}

Solving equation (\ref{equ}) under the above boundary conditions,
we obtain the displacement, $u(x,t)$, as  (See Appendix)
\begin{equation}
u(x,t)=\sum_n \frac{2v_{\infty} \tau}{n^2}\left[ (-1)^n \left( \cos{\frac{n\pi}{L}x} - 1 \right) \right]
\exp{\left(-\frac{n^2}{\tau}t\right)},
\label{displacement}
\end{equation}
where
\begin {eqnarray}
\tau &=& \frac {L^2 f}{\pi^2 \kappa},  \label{tau} \\
v_\infty &=& \frac {P_0}{Lf}.
\end {eqnarray}
The velocity of the water flow, $v(t)$, is derived from equation (\ref{eqv}) as 
\begin {equation}
v = v_\infty \left[ 1 + 2 \sum_{n=1}^{\infty} \exp \left( - \frac {n^2}{\tau} t \right) \right] = v_\infty \theta_3 (0, e^{t/\tau}),
\label{velocity}
\end {equation}
where $\theta_3(u,q)$ is the elliptic theta function.
The position of meniscus is expressed as
\begin {equation}
a \left[ t - 2 \sum_{n=1}^{\infty} \frac{\tau}{n^2} \exp \left( - \frac {n^2}{\tau} t \right) \right] + b,
\label{position}
\end {equation}
where $b$ is the initial position, and
\begin{equation}
a=\left( \frac{R}{r} \right)^2 v_\infty.
\end{equation}

\section{Discussion}

Now we compare the experimental results with the theoretical predictions
by equations (\ref{velocity}) and (\ref{position}).  The values obtained by
the experimental measurements are, $L = 1.0 \times 10^{-3}$ m, $\kappa
= 4.1 \times 10^{4}$ N/m$^{2}$, and $f = 6.6 \times 10^{14}$
Ns/m$^{4}$, respectively.  

By applying these values into
equation (\ref{tau}), we obtain the characteristic relaxation time
$\tau = 1.6 \times 10^{3}$ s.  The equilibrium velocity of water flow in
the gel observed by the experiments is, as is already given in the section \ref{sec:results}, $v_{\infty}
= 4.4 \times 10^{-9}$ m/s.   From equation (\ref{equilibrium}) the deformation of gel is estimated as
3.5$\%$ which is consistent with our assumption of small $u$.

The time evolution of the water flow is
then calculated by equation (\ref{velocity}) with above experimental values.  
The results are given in Figure 5.  
\begin{figure}
\resizebox{0.9\columnwidth}{!}{%
  \includegraphics{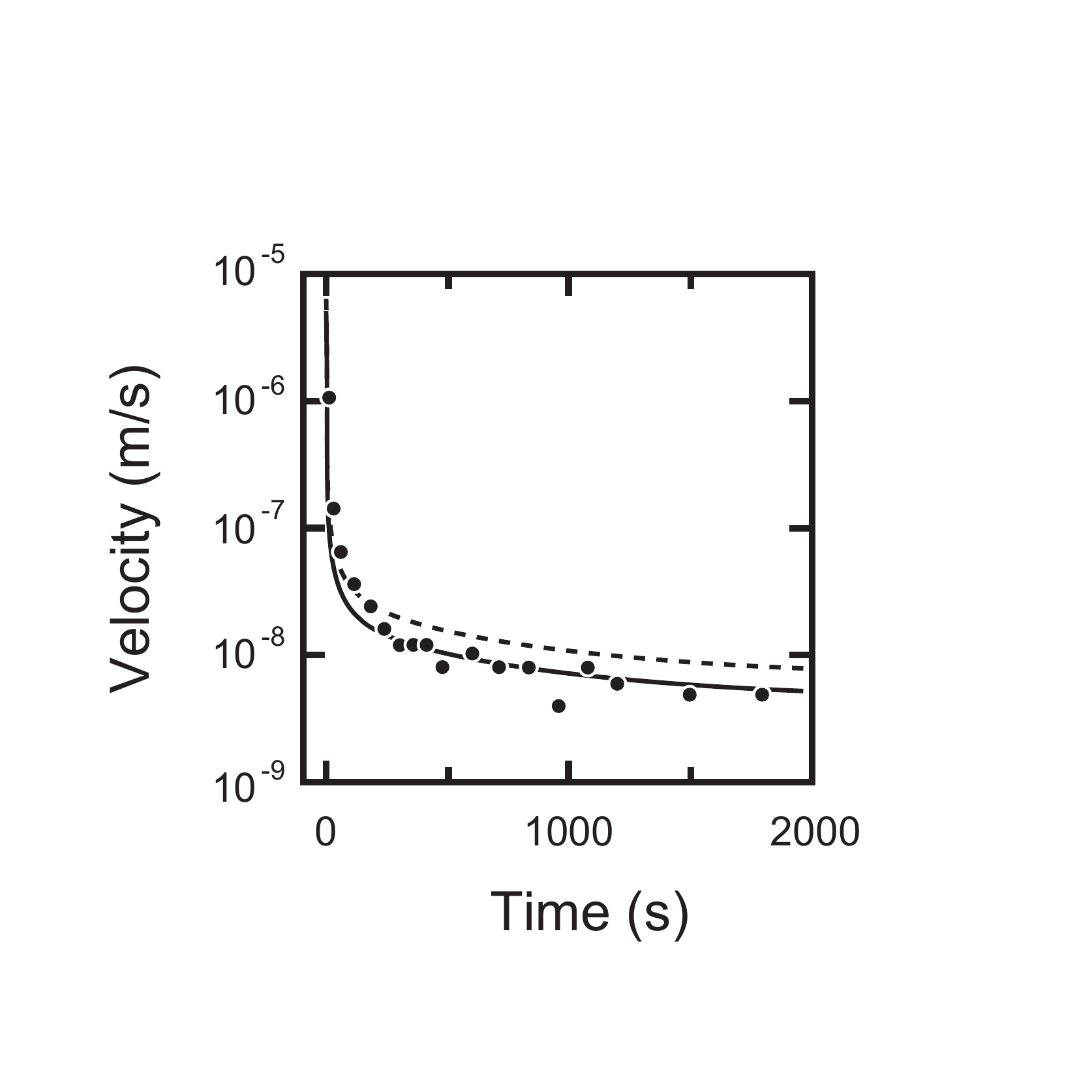}
}
\caption{Water flow velocity in gel: (a) the solid line corresponds to the theoretical curve fitting all the data, 
(b) the dotted line corresponds to the theoretical curve expected 
from the short time data ($<$200s).}
\label{fig:5}       
\end{figure}
The normalized velocity of the water flow in the gel is also calculated
from the slope of the flow curve in Figure 3 and shown in Figure 5.
We find that the time evolution of the velocity
of water flow in the gel is well explained by the equation (\ref{velocity}) 
with a single characteristic relaxation time $\tau$.  

However, there is a slight discrepancy 
in the time evolution curve of the velocity at a short time range ($<$200s) between
the theory and the experimental results as one can see in figure \ref{fig:3} and \ref{fig:5}.
The absolute value of the equilibrium velocity determined from the short time data ($<$200s) of the friction experiment is smaller than that determined by including the long time data.
Figure 5 shows that the deference between those two values is roughly 20 $\%$.  


It would be worth noting that the theory predicts that
the water flow creates a uni-axial deformation of polymer network,
i.e., the one dimensional concentration gradient in the gel,
proportional to
\begin{equation}
 \left( 1+ \frac{\partial u(x, \infty)}{\partial x} \right)^{-1} \approx 1+\frac{P_0 x}{D_c L f}. 
\end{equation}
The compression of gel is the cause of the
initial rapid water flow.
It is reported that the friction of the homegeneous gel depends on 
the polymer volume fraction, $\phi$, as $f \propto \phi^{1.5}$ \cite{TokitaJCP95,deGennes}. 
The friction, therefore, is expected to increase according to the compression of the
gel by the water flow.   The increment of the friction may explain the difference between the short time data and long time data.
The difference between the estimated equilibrium water velocities corresponds to 14 $\%$ homogeneous compression of the
sample slab gel, which is larger than the deformation which the present theory estimates.  Unfortunately, we cannot measure the deformation of the gel under the water flow
with the current experimental setup.

\section{Conclusion}
The kinetics of the water solvent flow through
the polymer network of gel is measured experimentally and described by a simple 
phenomenological theory.  The experiment shows that the equilibrium state is
reached after long time and that the friction of the gel shows a large value. 
The time evolution of the velocity of the
water flow in the gel is calculated on the basis of the
collective diffusion model of the polymer network coupled with the water friction,
assuming both the water velocity and the friction coefficient between the polymer network and the water are
uniform in the gel.  
The theory reproduces the time evolution of the water flow of the experiments for thin dilute gel 
fairly well with a single characteristic relaxation time.
We are able to deduce the collective diffusion coefficient from the water flow experiment on gels. 

However there is a difference between experimental value of the equilibrium velocity and
the value expected from the theoretical calculation by using parameters obtained 
from the short time flow measurements and the mechanical measurements.  
It suggests that the inhomogeneity of the friction and the concentration of the gel
under the water flow might be important for the kinetics of the water flow.
A better microscopic model is required for complete understanding of 
water flow through gel beyond the phenomenological understanding presented in this paper.

\begin{acknowledgement}
This work was first suggested by the late Professor Toyoichi Tanaka in 1989 
during the stay of Y.Y.S. and M.T. in Massachusetts Institute of Technology.
Y.Y.S. thanks CEA, IPhT and Takushoku University, RISE for the financial support.
M.T. thanks NEDO for the financial support.   
\end{acknowledgement}

%
%

\appendix

\section*{APPENDIX A: Solution of the equation of motion}

We define a function $W(x, t)$ as a deviation from the equilibrium displacement:
\begin {equation}
u(x, t) -\int v(t) dt \equiv W(x,t) + u(x, \infty).
\end {equation}
We obtain an equation for $W(x,t)$ from equation (\ref{equ}) as
\begin {equation}
\frac {\partial W}{\partial t} = D_c \frac {\partial^2 W}{\partial x^2}.
\label{eqw}
\end {equation}
By assuming $W(x,t) = X(x)T(t)$, the equation (\ref{eqw}) becomes
\begin {eqnarray}
\dot T(t) &=& - \lambda ^2 D_{c}T(t), \\
\frac {d^2 X(x)}{dx^2} &=& - \lambda^2 X(x),
\end {eqnarray}
where $\lambda$ is a constant. 
The general solution for $W(x,t)$ would be
\begin {eqnarray}
W(x,t) &=& \sum_{n=1}^\infty \left[ A_n \cos \lambda_n x  + B_n \sin \lambda_n x \right]   \nonumber \\
&& \times \exp (- \lambda_n^2 D_c t)+A_0 x+B_0.
\end {eqnarray}

The boundary conditions for $W(x,t)$ are
\begin {eqnarray}
\left. \frac {du}{dx} \right|_{x=0} = 0 &\longrightarrow& \left. \frac{dW}{dx}\right|_{x=0} = 0, \label{bcx=0}\\
\left. \frac {du}{dt}\right|_{t=\infty} = 0 &\longrightarrow& \left. \frac {dW}{dt}\right|_{t=\infty} = v(\infty), \label{bct}\\
u(L,t) = 0 &\longrightarrow& W(L, t) = - \int v dt. \label{bcL}
\end {eqnarray}
The boundary condition (\ref{bcx=0}) yields
\begin {eqnarray}
\left. \frac {dW}{dx} \right|_{x=0} 
&=&  \sum_{n=1}^\infty [ A_n (-\lambda_n \sin \lambda_nx) + B_n(\lambda_n \cos \lambda_n x)]  \nonumber \\
&& \left. \times \exp (- \lambda_n^2 D_c t) \right|_{x=0} + B_0 \nonumber \\
&=& \left. \sum_{n=1}^\infty [ A_n(- \lambda_n \times 0) + B_n (\lambda_n \times 1)] \exp (- \lambda_n^2 D_c t) \right|_{x=0} \nonumber \\
&&+B_0 \nonumber  \\
&=& 0. \nonumber 
\end {eqnarray}
This indicates $B_n =0$. 

The boundary condition (\ref{bct}) yields
\begin {eqnarray}
\left. \frac {dW}{dx} \right|_{t=\infty} &=& \left. \sum_{n=1}^\infty [A_n \cos \lambda_n x ](-\lambda_n^2 D_c) \exp (- \lambda_n^2 D_c t) \right|_{t=\infty} +A_0 \nonumber \\
&=& v(\infty).  \nonumber
\end {eqnarray}
This indicates that the constant, $\lambda_n^2$, should be positive, $\lambda_n^2 > 0$ for $n\geq 1$ and $A_0=v(\infty)$. 
Finally, the boundary condition (\ref{bcL}) yields
\begin {eqnarray}
W(L, t) &=& \sum_{n=1}^\infty [ A_n(\cos \lambda_nL + C_{nA})] \exp (- \lambda_n^2 D_c t) +A_0 L +B_0\nonumber \\
&=& - \int v dt.  \nonumber
\end {eqnarray}
This indicates $C_{nA} = - \cos \lambda_n L$.
The following relationships for $B_n$, $\lambda_n^2$, and $C_{nA}$ are obtained:
\begin {eqnarray}
B_n &=& 0, \\
\lambda_n^2 &>& 0, \\
C_{nA} &=& - \cos \lambda_n L.
\end {eqnarray}
And $W(x,t)$ becomes
\begin {equation}
 W(x,t) = \sum_{n=1}^{\infty} [A_n (\cos \lambda_n x + C_{nA})] \exp (-\lambda_n^2 D_c t).
\label{solw}
\end {equation}

Substitution of this solution (\ref{solw}) into equation (\ref{eqw}) yields
\begin {eqnarray}
& &\sum_{n=1}^{\infty} [A_n (\cos \lambda_n x + C_{nA})] (- \lambda_n^2 D_c) \exp (-\lambda_n^2 D_c t) \nonumber \\
& &- \frac {1}{L} \int_0^L  \sum_{n=1}^{\infty}  [A_n (\cos \lambda_n x + C_{nA})]   \nonumber \\
&& \times (- \lambda_n^2 D_c) \exp (-\lambda_n^2 D_c t)dx \nonumber \\
& & = D_c  \sum_{n=1}^\infty  A_n \lambda_n^2 (-\cos \lambda_n x) \exp (-\lambda_n^2 D_c t).
\label{eqw2}
\end {eqnarray}
The first term of the left handed member (LHM1) is written as follows:
\begin {eqnarray*}
{\rm LHM1} = && \sum_{n=1}^\infty (- \lambda_n^2 D_c)A_n (\cos \lambda_nx) \exp (- \lambda_n^2 D_c t)\\
&+& \sum_{n=1}^\infty (- \lambda_n^2 D_c) (A_n C_{nA}) \exp (- \lambda_n^2 D_c t).
\end {eqnarray*}
The second term of the left handed member (LHM2) is expressed as follows:
\begin {eqnarray*}
{\rm LHM2} = &-& \frac {1}{L} \int_0^L  \sum_{n=1}^{\infty} (- \lambda_n^2 D_c)  (A_n \cos \lambda_n x)   \\
&& \times \exp (-\lambda_n^2 D_c t)dx\\
&-& \frac {1}{L} \int_0^L  \sum_{n=1}^{\infty} (- \lambda_n^2 D_c)  (A_n C_{nA}) \exp (-\lambda_n^2 D_c t)dx.
\end {eqnarray*}
The first term of LHM2 is expressed as
\begin {eqnarray*}
&-& \frac {1}{L} \int_0^L  \sum_{n=1}^{\infty} (- \lambda_n^2 D_c)  (A_n \cos \lambda_n x) \exp (-\lambda_n^2 D_c t)dx\\
=  &-& \left. \frac {1}{L} \sum_{n=1}^{\infty} (- \lambda_n^2 D_c) \frac {1}{\lambda_n} (A_n \sin \lambda_n x) \exp (-\lambda_n^2 D_c t) \right|_0^L\\
=  &-& \frac {1}{L} \sum_{n=1}^{\infty} (- \lambda_n^2 D_c) \frac {1}{\lambda_n} (A_n \sin \lambda_n L) \exp (-\lambda_n^2 D_c t).
\end {eqnarray*}
The second term of LHM2 becomes
\begin {eqnarray*}
&-& \frac {1}{L} \int_0^L  \sum_{n=1}^{\infty} (- \lambda_n^2 D_c)  (A_n C_{nA}) \exp (-\lambda_n^2 D_c t)dx\\
= &-& \sum_{n=1}^{\infty} (- \lambda_n^2 D_c)  (A_n C_{nA}) \exp (-\lambda_n^2 D_c t).
\end {eqnarray*}
Hence, the equation (\ref{eqw2})  is reduced to
\[
- \frac {1}{L} \sum_{n=1}^\infty (- \lambda_n^2 D_c) A_n \frac {1}{\lambda_n} (\sin \lambda_n L) \exp (- \lambda_n^2 D_c t) = 0.
\]
This indicates
\begin {equation}
\lambda_n = \frac {n\pi}{L},
\end {equation}
and
\begin {equation}
C_{nA} = -\cos (n\pi) = -(-1)^n.
\end {equation}


The solution, $W(x,t)$, can be now written as
\begin {equation}
W(x,t) = \sum_{n=1}^\infty A_n \left(\cos \frac {n\pi}{L} x - (-1)^n \right) \exp \left( - \frac{n^2\pi^2}{L^2} D_c t \right).
\end {equation}
Since the initial condition is
\begin {equation}
W(x,0) = \frac {P_0}{2D_c Lf} (x^2 - L^2),
\end {equation}
the coefficients, $A_n$, are determined by the Fourier series expansion:
\begin {equation}
\sum_{n=1}^\infty A_n \left(\cos \frac {n\pi}{L} x - (-1)^n \right) = \frac {P_0}{2D_c Lf} (x^2 - L^2).
\end {equation}
We find
\begin {equation}
A_n = \frac {(-1)^n}{n^2 \pi^2} \frac {2P_0 L}{D_c f}.
\end {equation}

The displacement, $u(x,t)$, is now determined as
\begin {equation}
u(x,t) = \sum_{n=1}^{\infty} \mathcal {A}_n \left[ (-1)^n (\cos q_n x - 1) \right] \exp (-\Gamma_n t),
\end {equation}
where, $\mathcal {A}_n$, $q_n$, and $\Gamma_n$ are given by
\begin {eqnarray}
\mathcal {A}_n &=& \frac {2P_0 L}{n^2 \pi^2 D_c f}, \\
q_n &=& \frac {n\pi}{L}, \\
\Gamma_n &=& \left( \frac{n\pi}{L} \right)^2 D_c.
\end {eqnarray}


\begin{thebibliography}{}
%
%
\bibitem{TanakaSci}
T. Tanaka, Sci. Am. \textbf{244}, 124 (1981)
\bibitem{Muhr}
A.H. Muhr,  J.M.B. Blanshard, Polymer \textbf{23}, 1012 (1982)
\bibitem{TokitaPRE}
M. Tokita, T. Miyoshi, K. Takegoshi, K.Hikichi, Phys. Rev. E \textbf{53}, 1823 (1996)
\bibitem{TokitaJCP95}
M. Tokita, T. Tanaka, J. Chem. Phys. \textbf{95}, 4613 (1991)
\bibitem{TokitaSci}
M. Tokita, T. Tanaka, Science \textbf{253}, 1121 (1991)
\bibitem{TokitaAPS}
M. Tokita, Adv. Polym. Sci. \textbf{110}, 27 (1993)
\bibitem{TanakaPRL}
T. Tanaka, S. Ishiwata, C. Ishimoto, Phys. Rev. Lett. \textbf{38}, 771 (1977)
\bibitem{TanakaJCP70}
T. Tanaka, D.J. Fillmore, J. Chem. Phys. \textbf{70}, 1214 (1979)
\bibitem{Matsuo}
E.S. Matsuo, T. Tanaka, Nature \textbf{358}, 482 (1992)
\bibitem{TokitaJPSJ}
M. Tokita, S. Suzuki, K. Miyamoto, T. Komai, J. Phys. Soc. Jpn. \textbf{68}, 330 (1999)
\bibitem{MaskawaJCP}
J. I. Maskawa, T. Takeuchi, K. Maki, K. Tsujii, T. Tanaka, J. Chem. Phys. \textbf{105}, 1735 (1999)
\bibitem{TakigawaJCP111}
T. Takigawa, K. Uchida, K. Takahashi, T. Masuda, J. Chem. Phys. \textbf{111}, 2295 (1999)
\bibitem{TokitaJCP113}
M. Tokita, K. Miyamoto, T. Komai, J. Chem. Phys. \textbf{113}, 1647 (2000)
\bibitem{TakigawaJCP117}
T. Takigawa, T. Ikeda, Y. Takakura, T. Masuda, J. Chem. Phys. \textbf{117}, 17306 (2002)
\bibitem{BoudaoudPRE}
A. Boudaoud, S. Chaieb, Phys. Rev. E \textbf{68}, 021801 (2003)
\bibitem{UrayamaJCP}
L. Urayama, S. Okada, S. Nosaka, H. Watanabe, K. Takigawa, J. Chem. Phys. \textbf{122}, 024906 (2005)
\bibitem{NosakaPoly}
S. Nosaka, S. Okada, Y. Takayama, K. Urayama, H. Watanabe, T. Takigawa, Polymer \textbf{46}, 12607 (2005)
\bibitem{YoshikawaJJAP}
M. Yoshikawa, R. Ishii, J. Matsui, A. Suzuki, M. Tokita,  Jpn. J. Appl. Phys. Part 1 \textbf{44}, 8196 (2005)
\bibitem{SuzukiJCP}
A. Suzuki, M. Yoshikawa, J. Chem. Phys. \textbf{125}, 174901 (2006)
\bibitem{TanakaJCP59}
T. Tanaka, L.O. Hocker, G.B. Benedek, J. Chem. Phys. \textbf{59}, 5151 (1973)
\bibitem{deGennes}
P-G. de Gennes, \textit{Scaling Concepts in Polymer Physics} (Cornell University Press, Ithaca, London, 1979)
\bibitem{Munch1}
J.P. Munch, S. Candau, J. Hertz, G. Hild, J. Phys. (Paris) \textbf{38}, 971 (1977)
\bibitem{Munch2}
J.P. Munch, P. Lemarechal, S. Candau, J. Phys. (Paris) \textbf{38}, 1499 (1977)
\bibitem{TakigawaPJ1}
T. Takigawa, K. Urayama, and T. Masuda, Polymer J. \textbf{26}, 225 (1994)
\bibitem{TakigawaPJ2}
T. Takkigawa, Y. Morino, K. Urayama, and T. Masuda, Polymer J. \textbf{28}, 1012 (1996)
\bibitem{Yamaue}
T. Yamaue, M. Doi, Phys. Rev. \textbf{E69}, 041402 (2004)
\bibitem{Doi}
M. Doi, \textit{Dynamics and Patterns in Complex Fluids}, edited by A. Onuki and K. Kawasaki (Springer, New York, 1990), p.100
\end{thebibliography}
%

\end{document}